\documentclass
[showpacs,letterpaper,prl,amsfonts,amssymb,oneside,10pt,preprint,balancelastpage,notitlepage,twocolumn]{revtex4}%
\usepackage{amsfonts}
\usepackage{amsmath}
\usepackage{amssymb}
\usepackage{graphicx}%
\begin{document}
\title{Hysteretic nonequilibrium Ising--Bloch transition}
\author{Victor B. Taranenko, }
\affiliation{Institute of Physics, National Academy of Sciences of the Ukraine, Kiev, Ukraine}
\author{Adolfo Esteban--Mart\'{\i}n, Germ\'{a}n J. de Valc\'{a}rcel, and Eugenio Rold\'{a}n}
\affiliation{Departament d'\`{O}ptica, Universitat de Val\`{e}ncia, Dr. Moliner 50,
46100--Burjassot, Spain}

\begin{abstract}
We show that a parametrically driven cubic--quintic complex Ginzburg--Landau
equation exhibits a hysteretic nonequilibrium Ising--Bloch transition for
large enough quintic nonlinearity. These results help to understand the recent
experimental observation of this pheomenon [A. Esteban-Mart\'{\i}n \textit{et
al.}, Phys. Rev. Lett. \textbf{94}, 223903 (2005)].

\end{abstract}

\maketitle

Spatially extended bistable systems with broken phase invariance display
defects in the form of interfaces, so-called domail walls (DWs). A paradigm
for the study of DWs is the parametrically driven complex Ginzburg--Landau
equation, which can be written in the form \cite{Coullet}%
\begin{equation}
\partial_{t}A=\gamma A^{\ast}+\left(  \mu+i\nu\right)  A+\left(
1+i\alpha\right)  \partial_{x}^{2}A-\left(  1+i\beta\right)  \left\vert
A\right\vert ^{2}A, \label{PDNLSE}%
\end{equation}
where $\gamma$ is the parametric pump, $\mu$ accounts for linear gain or loss,
depending on its sign, $\nu$ is a detuning, $\alpha$ is the diffraction
coefficient, and $\beta$ is the nonlinear dispersion coefficient. In writing
Eq. (\ref{PDNLSE}) the spatial coordinate and the field amplitude have been
normalized to the square root of the diffusion coefficient and of the
saturation coefficient, respectively. This equation represents a universal
description of parametrically excited waves \cite{deValcarcel02} as well as of
the close to threshold dynamics of self-oscillatory systems externally forced
at the second harmonic of the natural oscillations frequency \cite{Coullet}.

In Eq. (\ref{PDNLSE}) the phase invariance of the field $A\left(  x,t\right)
$ is broken because of the presence of the parametric term $\gamma A^{\ast}$,
i.e, Eq. (\ref{PDNLSE}) shows the discrete symmetry $A\leftrightarrow-A$. This
makes possible the existence of domain walls (DWs) that connect spatial
regions where the field passes from, e.g., the homogeneous solution $A_{0}$ to
the equivalent solution $-A_{0}$. There are two types of DWs, namely Ising and
Bloch walls, which differ in the way the field crosses the complex zero at the
DW core: In the Ising wall, both the real and the imaginary parts of the field
become null, whilst in the Bloch wall the real and the imaginary parts become
null at different spatial points. Then, in terms of the field intensity,
$\left\vert A\right\vert ^{2}$, an Ising wall is dark at its center whilst it
is grey in the case of a Bloch wall. But the most striking difference between
Ising and Bloch walls lies in their different dynamic behaviour: When
nonvariational terms are present (in Eq. (\ref{PDNLSE}) this means that $\nu$,
$\alpha$ or $\beta$ be different from zero) Bloch walls move whilst Ising
walls remain at rest.

Coullet et al. \cite{Coullet} have discussed the above in detail for $\mu>0$
and have shown how Ising walls bifurcate into Bloch walls through the
so--called nonequilibrium Ising--Bloch transition (NIBT), that takes its name
from the equilibrium Ising--Bloch transition ocurring in ferromagnets
\cite{Bulaevskii}. Subsequently it was also shown that the NIBT occurs in Eq.
(\ref{PDNLSE}) for negative $\mu$ \cite{deValcarcel02}, in which case DWs
connect not only equivalent homogeneous solutions but also spatially modulated solutions.

There are a few experimental observations of this phenomenon. As far as we
know, it has been reported only in liquid crystals \cite{Meron} either
subjected to rotating magnetic fields, \cite{Frisch,Nasuno} or to an alternate
electrical voltage \cite{Kawagishi}. This last experiment constitutes a
particularly clear observation of the NIBT free from 2D effects, which
complicate front dynamics through curvature effects. We must add our very
recent observation of a \textit{hysteretic} Ising--Bloch transition in a
nonlinear optical cavity \cite{Esteban05}. This last experiment was carried
out in a photorefractive oscillator in a degenerate four--wave mixing
configuration \cite{Esteban05,Larionova04,Esteban04} and the cavity detuning
played the role of the control parameter. We found that for small positive
cavity detuning the system exhibits Ising walls. When detuning was increased,
Ising walls bifurcated into Bloch walls at a cavity detuning value $\nu_{IB}$,
and for $\nu>\nu_{IB}$ DWs were always of Bloch type. Then, when making a
reverse detuning scan we found that Bloch walls existed up to a detunig value
$\nu_{BI}$ where a new Ising--Bloch transition occurs. The interesting thing
is that $\nu_{BI}<\nu_{IB}$, and then there is a detuning domain, $\nu
_{BI}<\nu<\nu_{IB}$, where Ising and Bloch walls coexist.

The origin of the hysteresys was experimentally found to lie in the existence
of bistability in the homogeneous state of the system: The homogeneous state
exhibits bistability within a certain cavity detuning range between two
spatially homogeneous states, say $A_{\hom,1}$ and $A_{\hom,2}$, with
$\left\vert A_{\hom,1}\right\vert ^{2}>\left\vert A_{\hom,2}\right\vert ^{2}$.
It occurs to happen that DWs connecting $A_{\hom,1}$ with $-A_{\hom,1}$ are of
Ising type whilst those connecting $A_{\hom,2}$ with $-A_{\hom,2}$ are of
Bloch type. Eq. (\ref{PDNLSE}) does not give any insight into this type of
behaviour as the NIBT it exhibits is not hysteretic nor its homogeneous
solution exhibits bistability of the type we are describing. Here we try to
put some light into this problem by considering a straightforward
generalization of Eq. (\ref{PDNLSE}).

Hysteretic Ising--Bloch transitions have been theoretically described recently
in two very different contexts. On the one hand, it has been described in the
anisotropic XY--spin system in an oscillatory magnetic field \cite{Fujiwara}.
In that paper, Eq. (\ref{PDNLSE}) is studied for $v=\alpha$ $=\beta=0$ plus an
additional modulation term. As the system under study is variational, Bloch
walls do not move and the hysteresys is found on the average oscillation
period of the DW for certain parameter sets. On the other hand, hysteresis has
also been found in the routes leading from standing fronts to a couple of
counterpropagating ones in two discrete models (an array of Lorenz units and
the FitzHugh--Nagumo model) in Ref.\cite{Pazo}. These previous theoretical
results do not help to understand the experimental results in \cite{Esteban05}
that we have just resumed. As already outlined in \cite{Esteban05} we will
show that the addition of a quintic nonlinearity in Eq. (\ref{PDNLSE}) allows
to understand qualitatively the experimental results.

We start with the following natural generalization of Eq. (\ref{PDNLSE})%
\begin{align}
\partial_{t}A &  =\gamma A^{\ast}+\left(  \mu+i\nu\right)  A+\left(
1+i\alpha\right)  \partial_{x}^{2}A\nonumber\\
&  -\left(  1+i\beta_{3}\right)  \left\vert A\right\vert ^{2}A-\left(
\alpha_{5}+i\beta_{5}\right)  \left\vert A\right\vert ^{4}A,\label{quintic}%
\end{align}
i.e., we have added a fifth order nonlinearity with $\alpha_{5}$ and
$\beta_{5}$ the quintic saturation and nonlinear dispersion coefficientes,
respectively. As we are interested in the minimal modification of Eq.
(\ref{PDNLSE}) that contains a hysteretic NIBT, in the following we shall
concentrate on the special case $\alpha_{5}=0$, as the addition of the quintic
nonlinear dispersion term is enough for our purposes as we show next.

In Fig. 1 we represent the square modulus of the homogeneous solution of Eq.
(\ref{quintic}) as a function of detuning $\nu$ for $\gamma=2$, $\mu
=\alpha=-\beta_{3}=1$ and the different values of $\beta_{5}$ indicated in the
figure inset. Notice that there are two regions in which the homogeneous
solution is multivalued: For negative detuning, where there is coexistence
between two homogeneous solution values and the trivial solution, and also for
positive detuning whenever $\beta_{5}>\beta_{5}^{c}\approx0.39428$, where
there is coexistence of three homogeneous solutions. We find that the latter
requires that the signs of the nonlinear dispersion coefficients, $\beta_{3}$
and $\beta_{5}$ are different, and will concentrate on this case. Note that
Eq. (\ref{quintic}) holds the symmetry $\left(  A,\nu,\alpha,\beta_{3}%
,\beta_{5}\right)  \longleftrightarrow\left(  A^{\ast},-\nu,-\alpha,-\beta
_{3},-\beta_{5}\right)  $ and consequently, the behaviour of the homogeneous
solution is the same for the parameter sets $\left(  \nu,\beta_{3},\beta
_{5}\right)  $ and $\left(  -\nu,-\beta_{3},-\beta_{5}\right)  $.

The numerical integration of Eq. (\ref{quintic}) shows that for negative
detuning $\nu$, the system exhibits extended patterns and that the homogeneous
solution can be stable for positive $\nu$. Then it is for $\nu>0$ that we can
find DWs connecting homogeneous solutions and we concentrate in this case
(Notice that for $\beta_{5}=0$ this is the parameter region, $\nu>0$, where
the NIBT was studied in \cite{Coullet} and \cite{deValcarcel02}).

We have carried out the numerical integration of Eq. (\ref{quintic}) for
$\gamma=2$, $\mu=\alpha=1$ and different values of $\beta_{3}$ and $\beta_{5}%
$. We pass to comment first our results for $\beta_{3}=-1$ and different
values of $\beta_{5}$ and $\nu$.

In Fig. 2 we represent again the homogeneous steady state for the same
parameters as in Fig. 1 (except for the values of $\beta_{5}$), and have
marked the different patterns one can observe. For $\beta_{5}=0.38<\beta
_{5}^{c}$, Fig. 2(a), the homogeneous solution is single--valued and two types
of DWs are observed: Ising walls ($IW$ in the figure), for detunings $\nu
\leq1.24$; and Bloch walls ($BW$) for $\nu>1.24$. For $\nu>1.53$ the
homogeneous solution becomes modulationally unstable; Bloch walls connect
patterns in this region. In Fig. 3 both the intensity and phase spatial
profiles corresponding to an Ising wall, Fig. 3(a), and a Bloch wall, Fig.
3(b), are represented. Notice that the field intensity is null at the DW core
only in the Ising wall, and that the phase jump is sharp (smooth) for the
Ising (Bloch) wall. It is also interesting to notice the shoulder in the field
intensity in the case of the Bloch wall (the shoulder appears on the back side
of the wall with respect to the direction of movement).

As the value of $\beta_{5}$ is increased, new features appear. For $\beta
_{5}=0.395$ (i.e., slightly larger that $\beta_{5}^{c}$) the homogeneous
solution becomes multivalued, Fig. 2(b) (the dashed line indicates that the
homogeneous solution is unstable). In this case Ising walls do not bifurcate
directly into moving Bloch walls, but start oscillating periodically around a
fixed position, that is, there is not a net displacement of the wall. Fig. 4
shows the intensity and phase profiles of the oscillating wall in three
different instants of time and it can be appreciated how the wall pases from a
clear Bloch character (smooth phase jump) when it is at the center of the
oscillation, Fig. 4(b), to a clear Ising character (sharp phase jump) when it
is at the extremes of the oscillation, Figs. 4(a) and (c).

The behaviour just described appears when $\beta_{5}\simeq\beta_{5}^{c}$. When
$\beta_{5}$ is further increased a new and remarkabe effect appears: There is
a detuning range of coexistence between the oscillating and the Bloch walls,
see Fig. 2(c). This is better appreciated in Fig. 5 where the velocity of the
walls is represented as a function of detuning for the same parameters as in
Fig. 2. In Figs. 5(a) and 5(b) the behaviour of the velocity closely follows
that of the standard NIBT \cite{Coullet,deValcarcel02}, but in Fig. 5(c) the
new phenomenon of the hysteretic NIBT is clearly appreciated.

In the case we have described no coexistence between Ising and Bloch walls is
observed, only between oscillating and Bloch walls. But by decreasing the
value of $\beta_{3}$ from $-1$ to $-1.5$ we can observe this coexistence. In
Fig. 6 we represent again the wall velocity as a function of detuning for the
same values as in Fig. 2 except $\beta_{3}=-1.5$ and $\beta_{5}=0.6$. The
behaviour is similar to that described above but now there appears a wide
domain of coexistence between Bloch walls and both oscillating and Ising
walls. This is a much clearer hysteretic NIBT.

These results show that the addition of a quintic nonlinear dispersion term to
the parametrically driven complex Ginzburg--Landau equation suffices for
obtaining a hysteretic nonequilibrium Ising--Bloch transition (HNIBT), a
phenomenon first observed in \cite{Esteban05}. As a fifth-order nonlinearity
represents the simplest, higher order correction to the usual complex
Ginzburg--Landau, and the latter is of wide applicability in physical and
chemical systems, the HNIBT could be well observed in other systems. While we
do not claim that this simple model, Eq. (\ref{quintic}), represents an
accurate description of the experimental system in \cite{Esteban05} (the
photorefractive nonlinearity is saturating) we note that it describes
qualitatively the experimental observations, even reproducing small details
such as the shoulder in the field intensity in the case of the Bloch wall,
Fig. (3), that has been repeatidly observed during the experiments in
\cite{Esteban05}.

\begin{acknowledgments}
This work has been supported by Spanish Ministerio de Educaci\'{o}n y Ciencia
and European Union FEDER through projects BFM2002-04369-C04-01.
\end{acknowledgments}

{\LARGE Figure Captions}

\textbf{Fig. 1.} Homogeneous solution intensity as a function of detuning for
$\gamma=2$, $\mu=\alpha=-\beta_{3}=1$ and $\beta_{5}=0.35$ (i), $\beta
_{5}=0.3928$ (ii), and $\beta_{5}=0.45$ (iii), which is an enlargement of a
part of the positive detuning domain.

\textbf{Fig. 2. }Homogeneous solution intensity for the same parameter values
as Fig. 1 except $\beta_{5}$, which are marked in the figure. The different
pattern domains are marked as $IW$ (Ising walls), $BW$ (Bloch walls), $OW$
(oscillating walls), and $P$ (patterns). The continuous (dashed) line
indicates stable (unstable) homogeneous solution.

\textbf{Fig. 3. }Intensity (full line) and phase (dashed line) spatial
profiles of an Ising (a) and a Bloch (b) wall for the same parameters as Fig.
2(a) and the detuning values marked in the figure.

\textbf{Fig. 4. }Intensity (full line) and phase (dashed line) profiles of an
oscillating wall at three different instants of time (see text). The
parameters are the same as in Fig. 2(b) and $\nu=1.29$.

\textbf{Fig. 5. }Velocity of the domain walls as a funtion of detuning for the
same parameter values as in Fig. 2. The gray areas mark the velocitie of the
oscillating walls. The arrows mark the transition from Bloch walls to
oscillating walls when a decreasing detuning scan is carried out.

\textbf{Fig. 6. }Velocity of the domain walls as a function of detuning for
$\gamma=2$, $\mu=\alpha=1$, $\beta_{3}=-1.5$, and $\beta_{5}=0.6$.


\begin{thebibliography}{99}                                                                                               %


\bibitem {Coullet}P. Coullet, J. Lega, B. Houchmanzadeh and J. Lajzerowicz,
Phys. Rev. Lett. \textbf{65}, 1352 (1990).

\bibitem {deValcarcel02}G. J. de Valc\'{a}rcel, I. P\'{e}rez--Arjona and E.
Rold\'{a}n, Phys. Rev. Lett. \textbf{89}, 164101 (2002).

\bibitem {Bulaevskii}L. N. Bulaevskii and V. L. Ginzburg, Zh. Eksp. Teor. Fiz.
\textbf{45}, 772 (1963) [Sov. Phys. JETP \textbf{18}, 530 (1964)].

\bibitem {Meron}E. Meron, DDNS \textbf{4}, 217 (2000).

\bibitem {Frisch}T. Frisch, S. Rica, P. Coullet, and J. M. Gilli, Phys. Rev.
Lett. \textbf{72}, 1471 (1994).

\bibitem {Nasuno}S. Nasuno, N. Yoshimo, and S. Kai, Phys. Rev. E \textbf{51},
1598 (1995).

\bibitem {Kawagishi}T. Kawagishi, T. Mizuguchi, and M. Sano, Phys. Rev. Lett.
\textbf{75}, 3768 (1995).

\bibitem {Esteban05}A. Esteban--Mart\'{\i}n, V. B. Taranenko, J. Garc\'{\i}a,
G. J. de Valc\'{a}rcel, and E. Rold\'{a}n, Phys. Rev. Lett. \textbf{94},
223903 (2005).

\bibitem {Larionova04}Ye. Larionova, U. Peschel, A. Esteban-Mart\'{\i}n, J.
Garc\'{\i}a Monreal, and C. O. Weiss, Phys. Rev. A \textbf{69}, 033803 (2004).

\bibitem {Esteban04}A. Esteban-Mart\'{\i}n, J. Garc\'{\i}a, E. Rold\'{a}n, V.
B. Taranenko, G. J. de Valc\'{a}rcel, and C. O. Weiss, Phys. Rev A
\textbf{69}, 033816 (2004).

\bibitem {Fujiwara}N. Fujiwara, H. Tutu, and H. Fujisaka, Phys. Rev. E
\textbf{70}, 066132 (2004).

\bibitem {Pazo}D. Paz\'{o}, R.R. Deza, and V. P\'{e}rez--Mu\~{n}uzuri, Phys.
Lett. A \textbf{340}, 132 (2005).
\end{thebibliography}
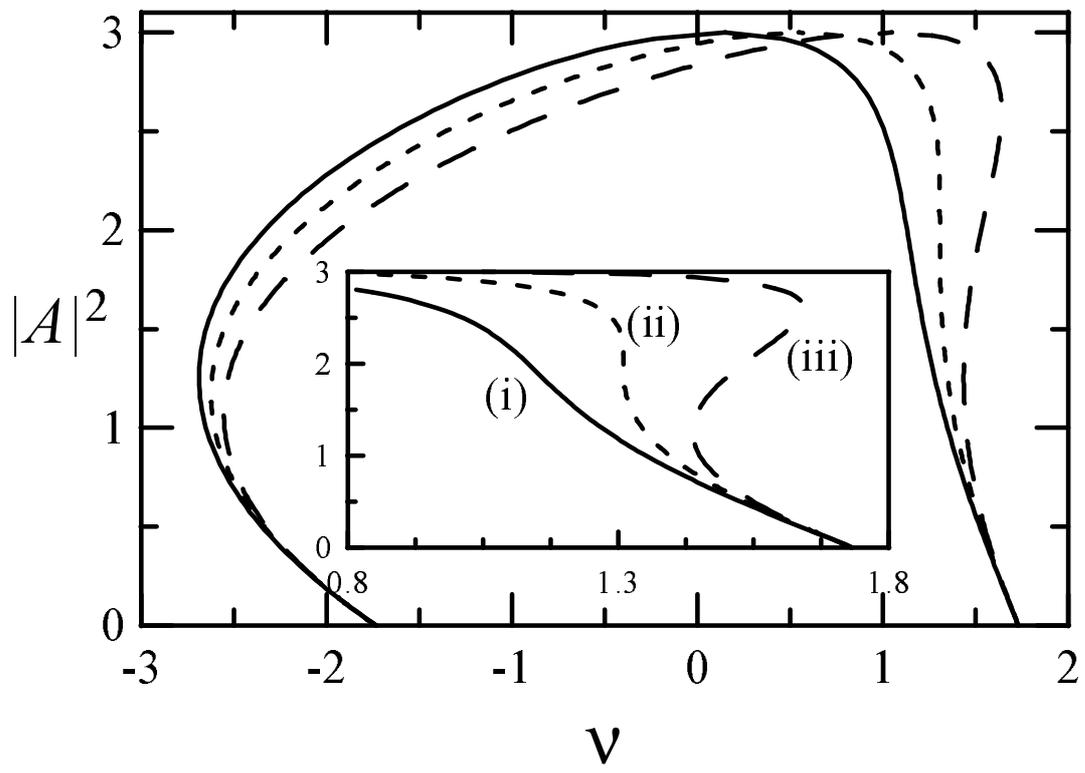
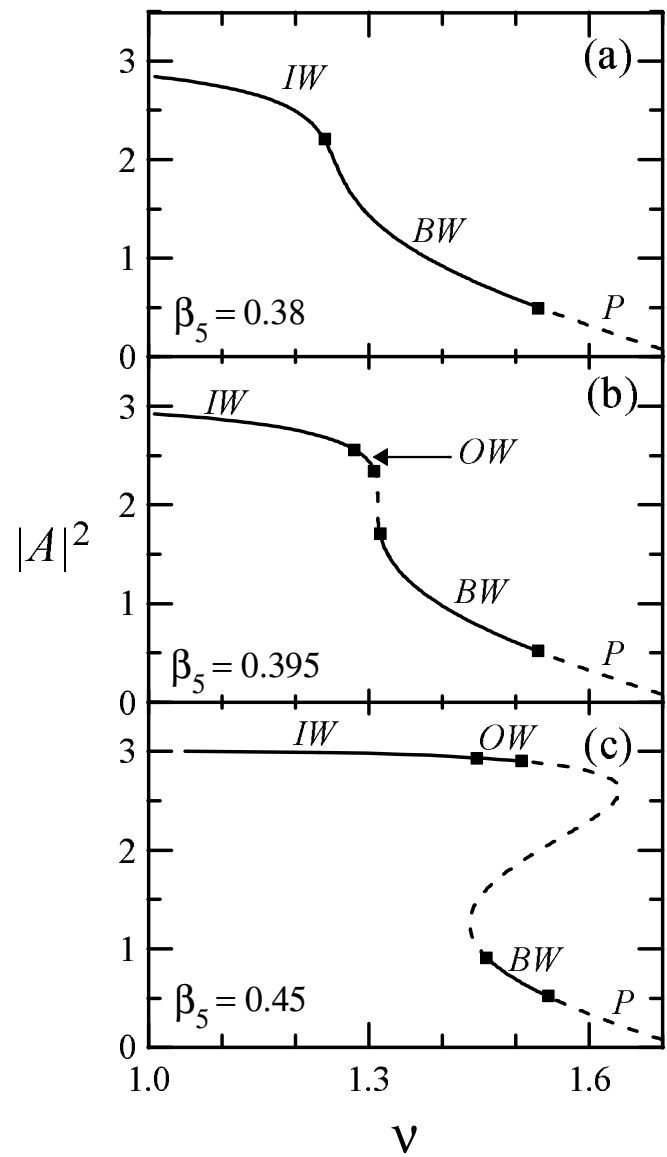
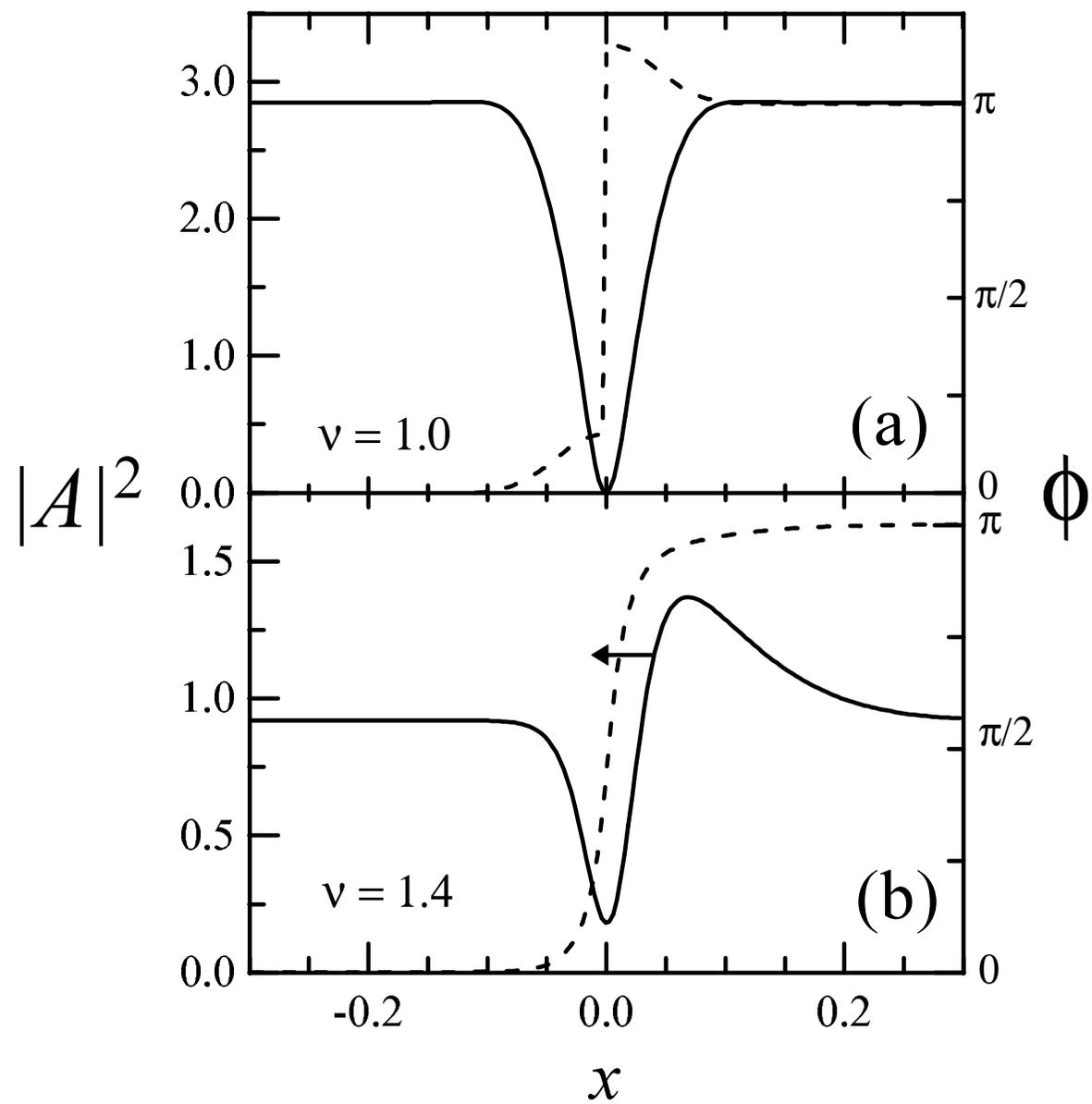
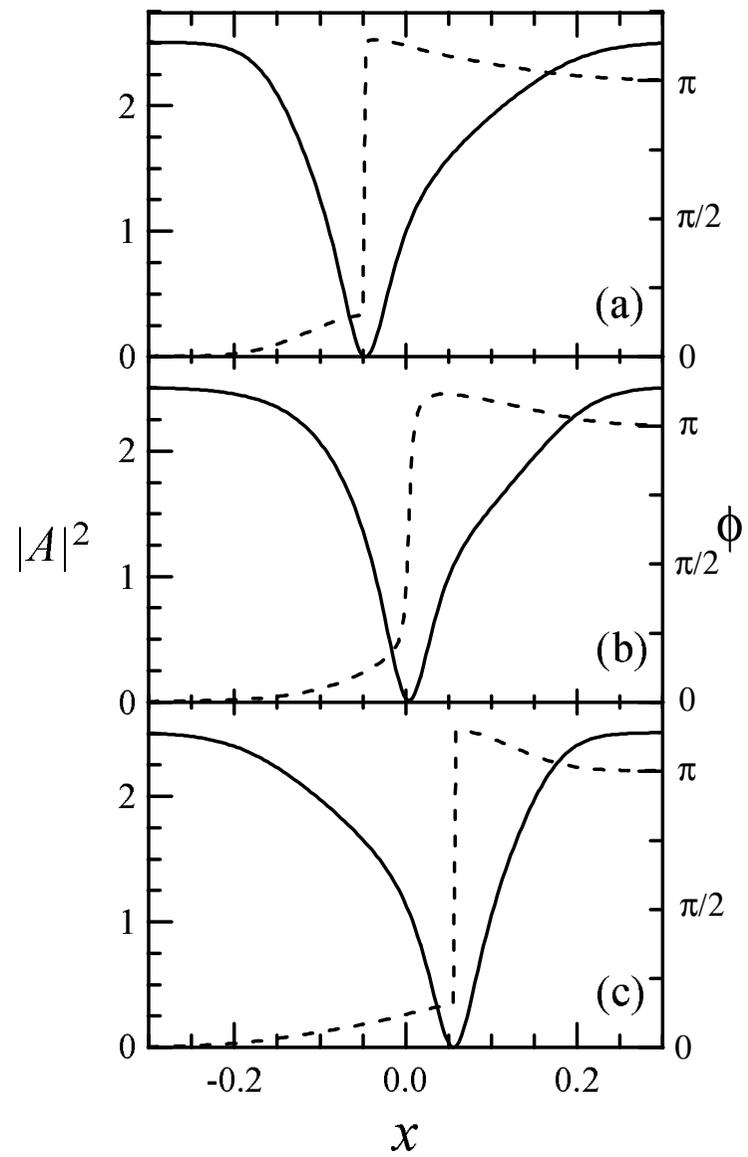
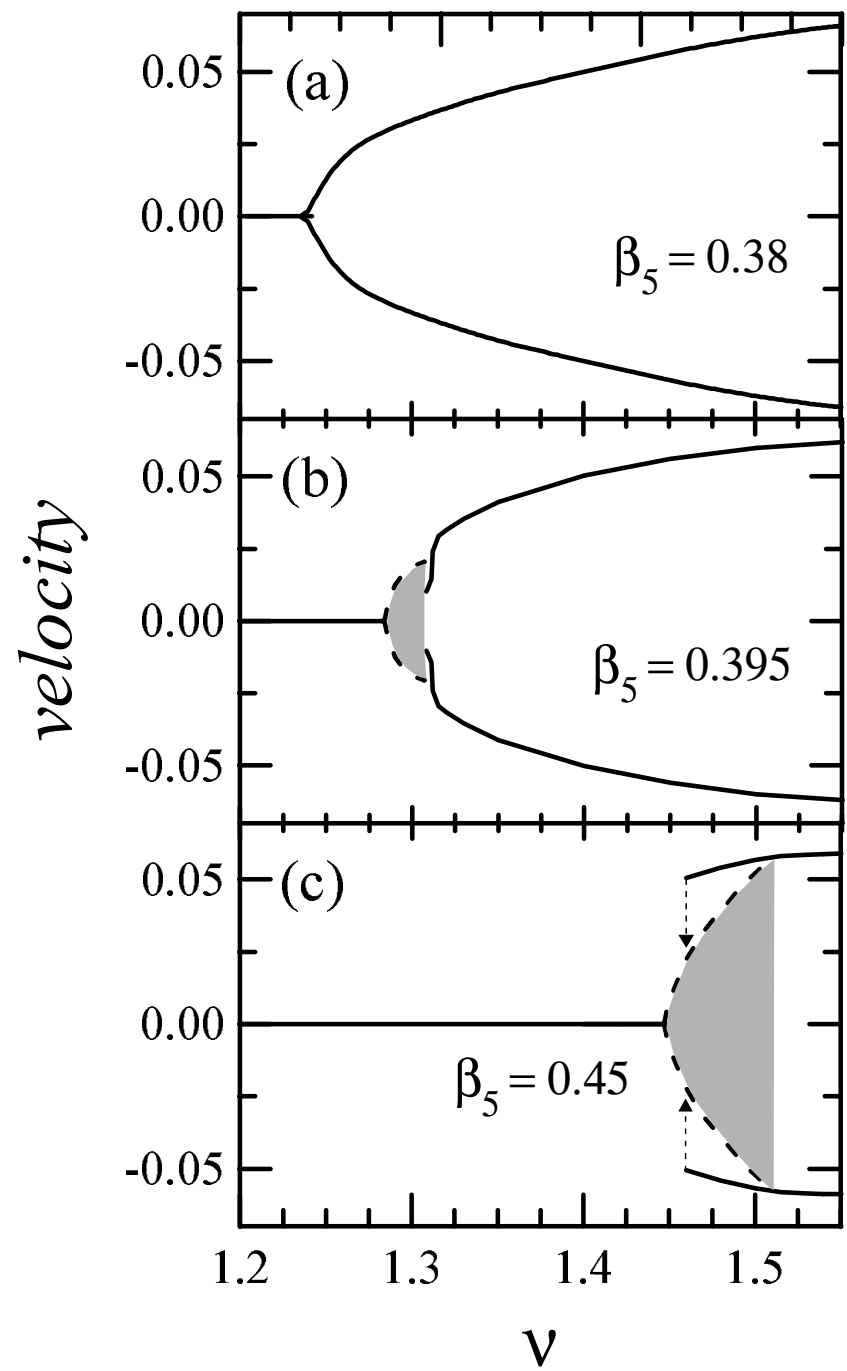
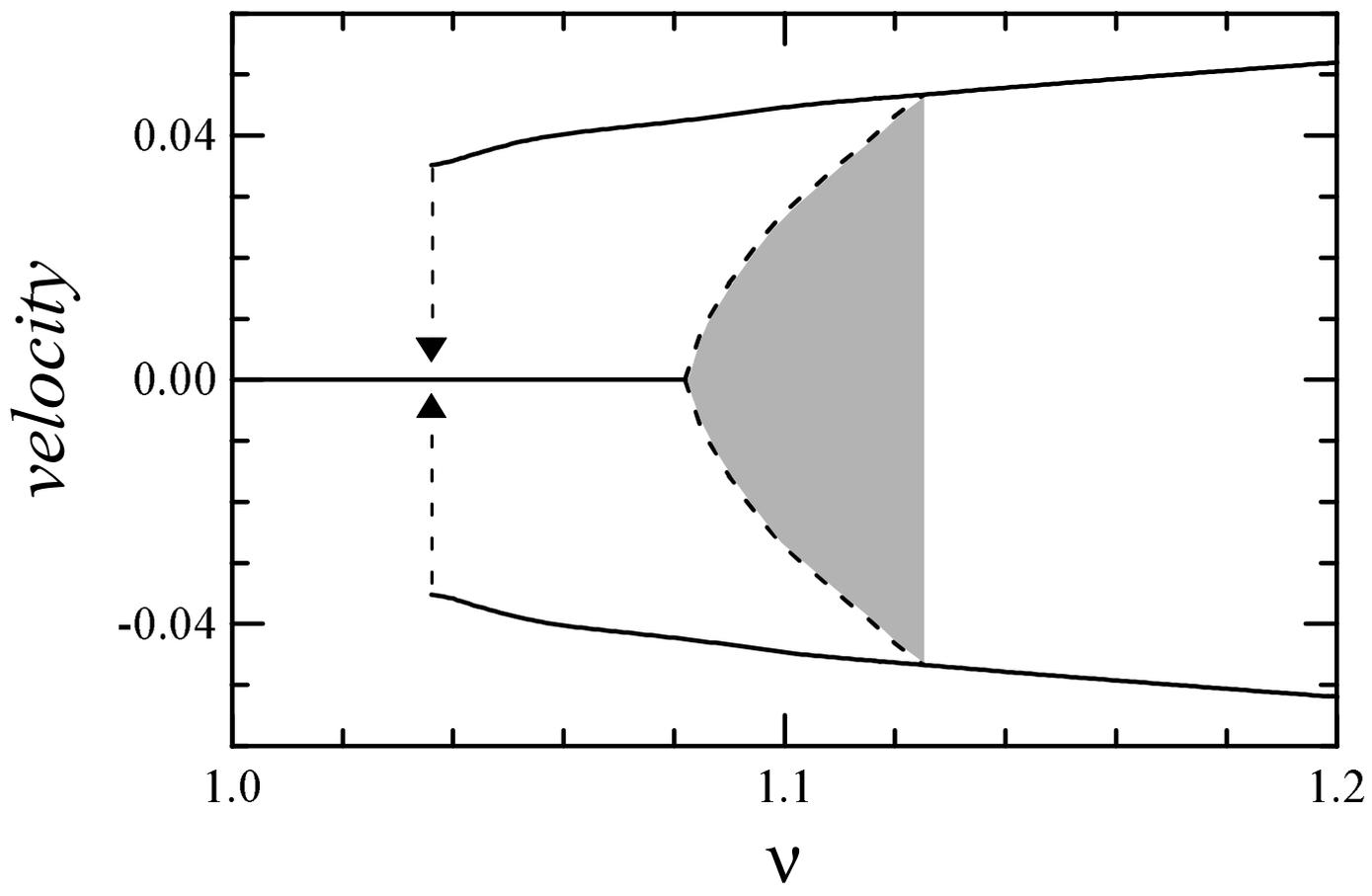
\end{document}